\documentstyle[11pt]{article}
\begin{document}
\begin{titlepage}
\begin{flushright}
YITP-00-27\\
June 2000
\end{flushright}
\begin{centering}
 
{\ }\vspace{1cm}
 
{\Large\bf On Electric Fields in Low Temperature Superconductors}\\

\vspace{2.0cm}

Jan Govaerts\footnote{On leave from the Institute of Nuclear
Physics, Catholic University of Louvain, Louvain-la-Neuve, Belgium\\
\indent{\ }\hspace{5pt}E-mail: {\tt govaerts@fynu.ucl.ac.be}}\\

\vspace{0.6cm}

{\em C.N. Yang Institute for Theoretical Physics}\\
{\em State University of New York at Stony Brook}\\
{\em Stony Brook NY 11794-3840, USA}\\

\vspace{0.6cm}

Damien Bertrand and Geoffrey Stenuit\footnote{
E-mail: {\tt bertrand, stenuit@fynu.ucl.ac.be}}\\

\vspace{0.6cm}

{\em Institut de Physique Nucl\'eaire, 
Universit\'e catholique de Louvain}\\
{\em 2, Chemin du Cyclotron, B-1348 Louvain-la-Neuve, Belgium}

\vspace{2cm}
\begin{abstract}

\noindent The manifestly Lorentz covariant Landau-Ginzburg equations
coupled to Maxwell's equations are considered as a possible framework 
for the effective description of the interactions between low temperature 
superconductors and magnetic as well as electric fields. A specific 
experimental set-up, involving a nanoscopic superconductor and only 
static applied fields whose geometry is crucial however, is described, 
which should allow to confirm or invalidate the covariant model through the
determination of the temperature dependency of the critical
magnetic-electric field phase diagram and the identification
of some distinctive features it should display.

\end{abstract}

\vspace{10pt}

\end{centering} 

\vspace{5pt}


\vspace{25pt}

\end{titlepage}

\setcounter{footnote}{0}

\noindent{\bf 1. Introduction.}
It is a widely held belief that (low $T_c$) superconductors cannot
sustain electric fields in static configurations. The argument\cite{Tinkham}
is directly based on the first of the London equations,
\begin{equation}
\frac{\partial}{\partial t}\left(\Lambda\vec{J}\right)=\vec{E}\ \ \ ,\ \ \ 
\vec{\partial}\times\left(\Lambda\vec{J}\right)=-\vec{B}\ \ \ ,
\label{eq:London}
\end{equation}
where $\Lambda$ is a phenomenological parameter proper to the superconducting
material, $\vec{J}$ is the supercurrent density, and $\vec{E}$, $\vec{B}$ are 
the electric and magnetic fields, respectively. Indeed, given that relation 
as well as the property of infinite conductivity, any nonvanishing electric 
field within the sample must set into motion a dissipationless supercurrent, 
hence a displacement of charges which very rapidly leads to an exact screening 
of any such electric field, certainly for time independent configurations 
(the case of stationary configurations with externally sustained supercurrents 
in the presence of magnetic vortices in type II superconductors is a different 
matter, see for example Refs.\cite{Josephson,Bok, Kolacek}).

However, such a situation raises a series of puzzles of varying degrees of
concern. First, it is physically inconceivable that an applied
electric field would discontinuously drop to zero from its external value
when crossing the surface of a superconducting sample. There
ought to exist some skin effect with a characteristic nonvanishing
penetration depth however small. Nevertheless, none of the parameters
appearing in the London equations provides for such an electric
penetration length. Moreover, in the above picture for the screening of
electric fields, mention is made of an electromagnetic
charge density which also is not accounted for in the London equations, 
nor more generally in the Landau-Ginzburg (LG) equations (see below).

Another concern at a more formal level is the fact that the coupling
of the London and LG equations to the electromagnetic fields is not
spacetime covariant. Indeed, under Lorentz boosts, the 
supercurrent density $\vec{J}$ ought to transform as the space
components of a 4-vector whose time component would then play
the role of the aforementioned missing charge density, while the electric 
$\vec{E}$ and magnetic $\vec{B}$ fields transform as components of the
two index antisymmetric field strength tensor $F_{\mu\nu}$.
To illustrate this point more vividly perhaps, consider
a flat infinite superconducting slab submitted to an external homogeneous
magnetic field lying parallel to its surface. In such a case,
the magnetic field will only partially penetrate the sample, with
a characteristic penetration length related to the parameter $\Lambda$
above\cite{Tinkham}. Imagine now performing a Lorentz boost in a 
direction both parallel to the surface of the slab and perpendicular 
to the applied magnetic field. According to the Lorentz covariance of 
Maxwell's equations, in the boosted frame there appears now an electric 
field perpendicular to the surface of the sample, {\sl also within 
the volume of the superconductor where the magnetic field in the 
initial frame is nonvanishing\/}. Thus Lorentz covariance requires
the possibility of electric fields on the same footing as magnetic
ones within superconductors, with an electric penetration length
equal to the familiar magnetic one. Nevertheless, the existence of 
such electric fields within superconductors is incompatible with 
the London equations.

One may take issue with the above covariance argument,
since the superconducting sample itself defines a preferred frame,
thereby insisting that the physics should be described only with respect 
to that specific rest-frame. Even though that frame is obviously 
distinguished, the coupling of the London and LG equations 
to the electromagnetic fields should be consistent with the covariance 
properties of Maxwell's equations, even within that frame. 
In addition, one would also like to have available
a manifestly covariant framework in which to study the interactions
of electromagnetic fields and moving superconducting samples,
an issue which the usual London and LG equations are unable to address,
and which is certainly accounted for to anyone's satisfaction in the 
description of electromagnetic interactions with ordinary conductors. It 
remains true nevertheless that some physical characterizations of 
superconductors can be defined with respect only to the rest-frame, such 
as for example the frame dependent notion of the free energy whose value 
determines the occurrence of the superconducting-normal phase transition
only when evaluated in the rest-frame.

Note that any spacetime covariant extension of the usual London and
LG equations entails a time dependent LG (TDLG) equation in which space and
time variations are on the same footing. Namely, a covariant
TDLG equation is necessarily of second-order in time derivatives
as it is in space derivatives, in sharp contrast with the usual
non covariant first-order TDLG equations encountered in the 
literature\cite{Tinkham}. In particular, in a covariant setting, the 
time scale associated to time dependent fluctuations is then naturally 
set by the time it takes light to travel the distance of some mean value 
of the penetration and coherence lengths. In contradistinction for
first-order TDLG equations, this time scale is specified
in terms of an additional parameter, the relaxation constant\cite{Tinkham}.
As a matter of fact, the latter quantity involves Boltzmann's constant,
showing that the usual TDLG equations apply rather to the time
dependency encured through thermodynamic fluctuations in the
superconducting order parameter. As such, these TDLG equations do not
provide a framework in which to study the intrinsically genuine time
dependent dynamics of superconductors coupled to time varying electromagnetic
fields, in the absence of thermodynamic fluctuations. A covariant
extension of the London and LG equations would also provide such
a dynamic framework, of relevance for instance to the
dynamics of ensembles of magnetic vortices interacting among one another
and with their electromagnetic environment.

A manifestly covariant extension of the LG equation which immediately
comes to one's mind is of course the so-called U(1) Higgs model of
particle physics, whose construction itself was motivated by the
BCS and LG theories in the late 50's. Indeed, as an effective theory
for superconductivity, this model coincides with the original LG
formulation for stationnary configurations, and readily provides\cite{Weinberg}
a description of all the remarkable quantum phenomena of superconductivity.
It may thus appear somewhat surprising that the covariant formulation
has not been used to further explore superconducting phenomena
possibly lying beyond the boundaries of the usual London and LG equations.
The purpose of this Letter is to suggest examples of
investigations along such lines.

After some considerations on the covariant London and LG equations presented
in the next section, section 3 identifies a specific set-up which should 
enable to establish experimentally whether the covariant, rather than 
the usual noncovariant approach is relevant to superconducting phenomena
in the presence of electric fields. What appears to us to be quite a
remarkable circumstance is that this experimental confirmation of the
covariant model should prove to be possible already using only a static, 
time independent configuration of external fields, through the observation 
of the superconducting-normal phase transition and the determination
of the phase diagram in the $(B,E)$ plane for a specific geometry of the 
applied fields.

\vspace{10pt}

\noindent{\bf 2. The covariant LG equations.}
As mentioned already, the considered model, of application only to low
temperature superconductors, is that of a U(1) gauge invariant coupling to 
the electromagnetic interactions of a complex scalar field $\psi$ of charge
$q=-2e<0$ (the Cooper pair charge) and with self-interactions determined
through the usual LG potential $(|\psi|^2-1)^2$ properly normalized. 
The order parameter $\psi$ is normalized to the square-root of the
Cooper pair density in the bulk in the absence
of any electromagnetic field (see Ref.\cite{Gov1} for some further
details). Rather than listing all the relevant equations
in terms of the physical quantities, let us already use the following
choice of units. Space and time coordinates, namely $\vec{x}$ and
the combination $x^0=ct$ with $c$ being the speed of light in vacuum,
are measured in units of the penetration length $\lambda(T)$. Similarly,
magnetic $\vec{B}$ and electric $\vec{E}/c$ fields are measured in units
of $\Phi_0/(2\pi\lambda^2(T))$, where $\Phi_0=2\pi\hbar/|q|$ is the usual
quantum of flux. Note that these units are temperature dependent,
since the penetration length $\lambda(T)$ is, with a dependency
we shall model\cite{Tinkham} through 
$\lambda(T)=\lambda(0)\left(1-(T/T_c)^4\right)^{-1/2}$, 
$T_c$ being of course the critical temperature. Finally, the order
parameter $\psi$ is parametrized according to $\psi=f e^{i\theta}$,
with $f$ real and $f^2$ thus measuring the relative Cooper pair density.
In terms of these units, space and time coordinates are denoted
$\vec{u}$ and $\tau$, and the magnetic and electric fields $\vec{b}$
and $\vec{e}$, respectively. Finally, let us also introduce the
quantities,
\begin{equation}
j^0=\frac{q}{\hbar}\frac{\lambda^3(T)}{f^2}\,\mu_0\,c\rho_{\rm em}\ \ \ ,\ \ \ 
\vec{j}=\frac{q}{\hbar}\frac{\lambda^3(T)}{f^2}\,\mu_0\,\vec{J}_{\rm em}\ ,
\end{equation}
where $\mu_0$ is the usual vacuum magnetic permitivity, and
$(c\rho_{\rm em},\vec{J}_{\rm em})$ are the superconducting electromagnetic
charge and current densities (constructed in terms of $\psi$), indeed
defining a 4-vector under Lorentz transformations.
Note that these relations show that $(f^2j^0,f^2\vec{j})$ is in fact
proportional to this electromagnetic 4-supercurrent, which must be 
locally conserved.

The latter remark is also confirmed by the inhomogeneous Maxwell
equations (all space derivatives are of course with respect
to $\vec{u}$),
\begin{equation}
\vec{\partial}\cdot\vec{e}=-f^2\,j^0\ \ \ ,\ \ \ 
\vec{\partial}\times\vec{b}-\partial_\tau\vec{e}=-f^2\vec{j}\ ,
\label{eq:Maxi}
\end{equation}
which indeed, as is usual, imply the local conservation property
$\partial_\tau\left(f^2j^0\right)+\vec{\partial}\cdot
\left(f^2\vec{j}\right)=0$.
The remaining electromagnetic equations of motion are given by,
\begin{equation}
\partial_\tau\vec{j}\,+\,\vec{\partial}j^0=-\vec{e}\ \ \ ,\ \ \ 
\vec{\partial}\times\vec{j}=\vec{b}\ ,
\label{eq:London2}
\end{equation}
which are recognized as the appropriate covariant extension of the 
London equations in (\ref{eq:London}). Note that only the first London 
equation is modified by the inclusion of a contribution of the supercharge
density, as was indeed required, while the second London equation, 
essential to the Meissner effect, retains its original form.
The homogeneous Maxwell equations,
\begin{equation}
\vec{\partial}\times\vec{e}+\partial_\tau\vec{b}=\vec{0}\ \ \ ,\ \ \ 
\vec{\partial}\cdot\vec{b}=0\ ,
\end{equation}
follow from the covariant London equations (\ref{eq:London2}) (as they
do also from the noncovariant ones in (\ref{eq:London})).

The equations of motion for the order parameter $\psi$ are given,
on the one hand, by the covariant LG equation
\begin{equation}
\left[\vec{\partial}^2-\partial^2_\tau\right]f=
\left[\vec{j}^2-{j^0}^2\right]f-\kappa^2(1-f^2)f\ ,
\label{eq:LG}
\end{equation}
where the LG parameter $\kappa=\lambda(T)/\xi(T)$---$\xi(T)$ being
the coherence length---is essentially temperature independent\cite{Tinkham},
and on the other hand, by the following conditions for the quantum phase 
$\theta$,
\begin{equation}
\partial_\tau\theta=-j^0+\varphi\ \ \ ,\ \ \ 
\vec{\partial}\theta=\vec{j}-\vec{a}\ .
\label{eq:theta}
\end{equation}
In these latter relations, $\varphi$ and $\vec{a}$ are the scalar and
vector gauge potentials defined such that
\begin{equation}
\vec{e}=-\vec{\partial}\varphi-\partial_\tau\vec{a}\ \ \ ,\ \ \ 
\vec{b}=\vec{\partial}\times\vec{a}\ .
\end{equation}
Finally, these equations are subject to boundary conditions requiring
vanishing values for those components of the vectors $\vec{\partial}f$ 
and $\vec{j}$ which are perpendicular to the surfaces of the 
superconducting sample in contact with an insulating material. 

The advantage of using this representation of the system is that the
number of gauge dependent variables is kept to a minimum\cite{Gov1}. Only the
quantities $\theta$, $\varphi$ and $\vec{a}$ are defined up to the following
gauge transformations,
\begin{equation}
\theta'=\theta+\chi\ \ ,\ \ 
\varphi'=\varphi+\partial_\tau\chi\ \ ,\ \ 
\vec{a}'=\vec{a}-\vec{\partial}\chi\ ,
\end{equation}
$\chi(\vec{u},\tau)$ being an arbitrary function, while the absolute
sign of $f$ may also be subject to gauge transformations\cite{Gov2}.

In fact, this decoupling of gauge variant and gauge invariant quantities
may be rendered complete when substituting the covariant London equations
(\ref{eq:London2}) into the inhomogeneous Maxwell equations (\ref{eq:Maxi}).
In addition to the covariant LG equation (\ref{eq:LG}), one then finds
for the $(j^0,\vec{j})$ 4-supercurrent,
\begin{equation}
\left[\vec{\partial}^2-\partial^2_\tau\right]j^0=
f^2 j^0-\partial_\tau
\left[\partial_\tau j^0+\vec{\partial}\cdot\vec{j}\right]\ \ \ ,\ \ \ 
\left[\vec{\partial}^2-\partial^2_\tau\right]\vec{j}=
f^2\vec{j}+\vec{\partial}
\left[\partial_\tau j^0+\vec{\partial}\cdot\vec{j}\right]\ .
\label{eq:current}
\end{equation}
Any solution to this set of three coupled equations for $j^0$, $\vec{j}$
and $f$ then leads to specific values for $\vec{b}$ and $\vec{e}$ 
through the covariant London equations (\ref{eq:London2}), and in turn, 
once the gauge potentials $\varphi$ and $\vec{a}$ related to these fields 
determined up to the gauge transformations parametrized by $\chi$, 
the corresponding solution for the quantum phase $\theta$ is also 
finally obtained from (\ref{eq:theta})\cite{Gov2}.

Although nonlinear, the equations (\ref{eq:LG}) and (\ref{eq:current})
also establish that these covariant LG equation admit progressive
wave solutions with covariant dispersion relations in a linear
regime. As a matter of fact, the phase and group velocities for 
variations in the 4-supercurrent and the order parameter $f$ are different, 
unless the LG parameter $\kappa$ takes the same critical value 
$\kappa_c=1/\sqrt{2}$ as the one which is so crucial to the understanding
of the stability and interaction properties of magnetic vortices.
Indeed, for a fluctuation of wave number $k$ (in units of $1/|\vec{u}|$)
around the vacuum solution
$(j^0=0,\vec{j}=\vec{0},f=1)$, the group velocities are,
respectively,
\begin{equation}
v_j=\frac{k}{\sqrt{k^2+1}}\ \ \ ,\ \ \ 
v_f=\frac{k}{\sqrt{k^2+2\kappa^2}}\ ,
\end{equation}
thus showing that when $\kappa>\kappa_c$ (resp. $\kappa<\kappa_c$),
waves in the 4-supercurrent will overtake (resp. be overtaken by)
those in the order parameter $f$. In other words, within the superconductor,
fluctuations in the electromagnetic fields will propagate more rapidly 
(resp. slowly) than those in the order parameter, in accordance with 
the relative magnetic and superconducting rigidities that the parameters 
$\lambda(T)$ and $\xi(T)$ characterize.

Clearly, such properties are totally different in the case of the
usual noncovariant first-order TDLG equation, in which time scales are then 
normalized with respect to the relaxation parameter, which itself is
temperature dependent. The ensuing dispersion relations are then
linear in frequency, implying that the group velocities of fluctuations
in the supercurrent $\vec{j}$ and in the order parameter $f$ are
then also identical, independently of the value for $\kappa$.
Such differences between the covariant and noncovariant frameworks
must lead to distinct physical properties in the case of time dependent
configurations in ultra-high frequency regimes, 
$\nu\sim c/\lambda(T),c/\xi(T)$, an issue which, however, is beyond
the scope of this work.

To conclude this general discussion, let us
also give the expression for the free energy $E$ of the system in 
the covariant form,
\begin{displaymath}
\left(\frac{\lambda^3(T)}{2\mu_0}
\left(\frac{\Phi_0}{2\pi\lambda^2(T)}\right)^2\right)^{-1}\,E=
\int_{(\infty)}d^3\vec{u}\left\{
\left[\vec{e}-\vec{e}_{\rm ext}\right]^2+
\left[\vec{b}-\vec{b}_{\rm ext}\right]^2\right\}\ +\
\end{displaymath}
\begin{equation}
+\ \int_\Omega d^3\vec{u}\left\{
\left(\partial_\tau f\right)^2+\left(\vec{\partial}f\right)^2+
f^2\left({j^0}^2+\vec{j}^2\right)+\frac{1}{2}\kappa^2\left(1-f^2\right)^2
-\frac{1}{2}\kappa^2\right\}\ ,
\label{eq:E1}
\end{equation}
where the normalization factor related to our choice of units is displayed
together with $E$ in the l.h.s., $\vec{e}_{\rm ext}$ and $\vec{b}_{\rm ext}$ 
are externally applied electric and magnetic fields, respectively, and
$\Omega$ stands for the volume of the superconducting sample.

The same expression is also of application to the noncovariant model,
in which case one has $j^0=0$ and $\vec{e}=\vec{0}$ within the superconductor,
and the quadratic term in $\partial_\tau f$ is to be replaced by 
a linear term while the time coordinate is then also measured in units of
the relaxation parameter for the TDLG equation. The term in
$\left[\vec{e}-\vec{e}_{\rm ext}\right]^2$ measures the energy required
to expulse the electric field from the superconductor. In the
noncovariant case, and in accordance with the first London equation,
we shall thus assume that the associated penetration depth
is essentially vanishing for all practical purposes. For physics
reasons, such an approximation cannot be very reliable when it
comes to nanoscopic superconductors, but we shall use it as
a working hypothesis anyway. Note that the free energy $E$
is defined here in such a way that it vanishes at the superconducting-normal
phase transition. And as a last remark, clearly, in the case of
stationary configurations and in the absence of any electric fields, 
the equations of both approaches coincide
with the usual noncovariant time independent LG equations.

\vspace{10pt}

\noindent{\bf 3. Characterizing the phase transition.}
In order to identify a specific geometry of applied fields which could 
help discriminate experimentally between the two approaches already in a
static configuration, consider again the situation of the flat infinite
slab of the Introduction, this time subjected not only to the homogeneous
magnetic field parallel to its surface, but also to an homogeneous
electric field applied perpendicularly to its surface (an electric 
field parallel to the slab does not induce a supercharge
distribution $j^0$, hence neither a feature distinctive from the 
noncovariant model). The slab is
taken to be of thickness $2a$, while the external electric field 
$\vec{e}_{\rm ext}$ is aligned along the $x$ axis, and the external magnetic 
field $\vec{b}_{\rm ext}$ along the $y$ axis, with components
$e_{\rm ext}$ and $b_{\rm ext}$, respectively (the origin of this
coordinate system is of course positioned in the center of the slab). 
For this specific geometry, the expulsion of the magnetic field is
achieved through an induced supercurrent $\vec{j}$ circulating along
the $z$ axis, also parallel to the slab, while that of the electric field
is achieved through the appearance of a nonvanishing supercharge density 
$j^0$, an occurrence which simply cannot arise in the noncovariant approach. 
Both these effects imply a deviation from its canonical value of unity
for the order parameter $f$. In view of the symmetries of the problem, both
$j^0(u)$ and $j^z(u)$ are odd functions of the normalized $u=x/\lambda(T)$ 
coordinate along the $x$ axis, while $f(u)$ is even.

Given the equations and the different conditions imposed on these quantities
at the boundaries $u=\pm u_a\equiv\pm a/\lambda(T)$, it proves possible as
well as useful to express both $j^0(u)$ and $j^z(u)$ in terms of a single
function $j(u)$
\begin{equation}
j^0(u)=-e_{\rm ext} j(u)\ \ \ ,\ \ \ 
j^z(u)=-b_{\rm ext} j(u)\ ,
\end{equation}
so that one has for the electric and magnetic fields within the
sample,
\begin{equation}
e(u)=e_{\rm ext}\frac{d}{du}j(u)\ \ \ ,\ \ \ 
b(u)=b_{\rm ext}\frac{d}{du}j(u)\ ,
\end{equation}
thus showing once again that Lorentz covariance implies that the
penetration lengths for both types of fields are identical.
The set of equations to be considered then reduces to,
\begin{equation}
\frac{d^2}{du^2}j(u)=f^2(u)j(u)\ \ \ ,\ \ \ 
\frac{d^2}{du^2}f(u)=\left(b^2_{\rm ext}-e^2_{\rm ext}\right)j^2(u)f(u)-
\kappa^2\left(1-f^2(u)\right)f(u)\ ,
\label{eq:slab1}
\end{equation}
subject to the boundary conditions
\begin{equation}
\frac{d}{du}j(u)_{|u=\pm u_a}=1\ \ \ ,\ \ \ 
\frac{d}{du}f(u)_{|u=\pm u_a}=0\ . 
\end{equation}
Note already the subtle interplay between the magnetic and electric
field contributions to the LG equation for $f(u)$, which leads to
values larger than unity for $f(u)$ in the electric regime
$e^2_{\rm ext}>b^2_{\rm ext}$, while, as is usual, $f(u)$ remains less
than unity in the magnetic regime $b^2_{\rm ext}>e^2_{\rm ext}$.
The existence of these two regimes is a direct and distinctive
consequence of manifest Lorentz covariance; only the magnetic one arises
in the noncovariant approach (see below). Note also that in view
of these equations, the solutions for $j(u)$ and $f(u)$ are necessarily
functions of the specific combination $(b^2_{\rm ext}-e^2_{\rm ext})$ only,
indeed justifying this notion of electric or magnetic regimes.

Up to the normalisation factor displayed in (\ref{eq:E1}) as well
as the infinite surface of the slab, the free energy ${\cal E}$
of configurations obeying these equations is simply given by
\begin{equation}
{\cal E}=2u_a\left\{\left[1-\frac{1}{u_a}j(u_a)\right]
\left(b^2_{\rm ext}+e^2_{\rm ext}\right)\ -\
\frac{1}{u_a}\int_0^{u_a}du\left[
\left(b^2_{\rm ext}-e^2_{\rm ext}\right)j^2f^2+
\frac{1}{2}\kappa^2f^4\right]\right\}\ .
\end{equation}
Consequently, the curve in the $(b,e)$ phase diagram which 
characterizes the superconducting-normal phase transition obeys
the following equation in the covariant approach,
\begin{equation}
b^2+e^2=\frac{1}{\left[1-\frac{1}{u_a}j(u_a)\right]}\,
\frac{1}{u_a}\int_0^{u_a}du\left[
\left(b^2-e^2\right)j^2f^2+
\frac{1}{2}\kappa^2f^4\right]\ .
\label{eq:critical1}
\end{equation}
The noteworthy property of this relation is that the l.h.s. involves
only the combination $(b^2+e^2)$, while the r.h.s. is only a function of
the combination $(b^2-e^2)$ of the external fields (since the solutions
$j(u)$ and $f(u)$ also share that property).

Before addressing the specific consequences of this equation for the
$(b,e)$ phase diagram, let us consider the corresponding expressions in
the noncovariant approach. In that case, one has of course $j^0(u)=0$
(which also implies $e(u)=0$ within the superconductor, as follows
from the first London equation) as well as $j^z(u)=-b_{\rm ext}j(u)$.
The equations then remain as given in (\ref{eq:slab1}), including the
boundary conditions, with the only but important difference that the factor
$(b^2_{\rm ext}-e^2_{\rm ext})$ appearing in the LG equation is
of course replaced by $b^2_{\rm ext}$ only. Hence, the noncovariant
LG equations only admit the magnetic regime of solutions.
Note that in this case, the solutions for $j(u)$ and $f(u)$ are
then also functions of the $b^2_{\rm ext}$ external field only.
Consideration of the expression for the free energy then leads to
the following condition of criticality in the $(b,e)$ phase diagram
in the noncovariant case,
\begin{equation}
b^2+\frac{1}{\left[1-\frac{1}{u_a}j(u_a)\right]}e^2=
\frac{1}{\left[1-\frac{1}{u_a}j(u_a)\right]}\,
\frac{1}{u_a}\int_0^{u_a}du\left[
b^2j^2f^2+\frac{1}{2}\kappa^2f^4\right]\ .
\label{eq:critical2}
\end{equation}
In spite of the apparent similarity with (\ref{eq:critical1}), recall
however that the r.h.s. of this expression is a function of $b^2$ only,
while the l.h.s. is no longer the simple combination $b^2+e^2$
characteristic of a circle since the coefficient multiplying the term
in $e^2$ is also a function of $b^2$.

Note that the conditions (\ref{eq:critical1}) and (\ref{eq:critical2})
coincide in the limit that no electric field is applied, $e=0$, as they
should of course. Moreover in the absence of any magnetic field, $b=0$,
(\ref{eq:critical2}) implies the existence of a nonvanishing critical
electric field, $e_0=\kappa/\sqrt{2}$, or in physical units
$E_0(T)/c=\left(\lambda(0)/\lambda(T)\right)^2B^\infty_c(0)$,
$B^\infty_c(0)=\Phi_0/(2\sqrt{2}\pi\lambda(0)\xi(0))$ being the usual
thermodynamic critical magnetic field in the bulk at zero temperature.
Clearly, the existence of such a critical electric field even in the
noncovariant approach is consequence of our definition for the free
energy in (\ref{eq:E1}) which accounts for the expulsed electric
energy density through the term in $\left[\vec{e}-\vec{e}_{\rm ext}\right]^2$.
In particular, this critical electric field $E_0(T)$ vanishes
at the critical temperature $T_c$, as does the critical magnetic field
$B_0(T)$ in the absence of any electric field, $e=0$.

\vspace{10pt}

\noindent{\bf 4. The $(B,E)$ phase diagram.}
A complete unravelling of the consequences of the criticality conditions
(\ref{eq:critical1}) and (\ref{eq:critical2}) requires of course
a numerical approach. Nevertheless, an analysis in some limiting
situations already suffices to gain insight into the differences
implied by the two models. An obvious such situation is obtained in
the macroscopic limit, namely when the slab half-thickness $a$ is
much larger than both the penetration and coherence lengths. For all
practical purposes, the function $j(u)$ then essentially vanishes
whereas the order parameter retains its canonical value of unity 
within most of the volume of the sample, except for a small region close
to the surface. Hence in the above expressions of criticality,
in the limit that $a\rightarrow\infty$, only the contribution in
$\kappa^2f^4/2$ tends to dominate, leading in both cases to the
condition,
\begin{equation}
b^2+e^2\simeq \frac{1}{2}\kappa^2\ \ \ ,\ \ \ a\gg\lambda(T),\xi(T)\ .
\label{eq:macro1}
\end{equation}
Since this will prove to be useful, let us normalize the measurement
of these fields to the value $b_0$ of the critical magnetic
field in the absence of any electric field (in the macroscopic limit,
we thus have $b_0=\kappa/\sqrt{2}$). In terms of the physical
quantities, one then obtains the following approximation to the
criticality condition in the $(B,E)$ phase diagram
\begin{equation}
\left(\frac{B}{B_0}\right)^2+\left(\frac{E/c}{B_0}\right)^2\simeq 1\ \ \ ,
\ \ \ a\gg\lambda(T),\xi(T)\ .
\label{eq:macro2}
\end{equation}
Hence in the macroscopic limit, the two models are not distinguished
in their $(B,E)$ phase diagrams. In particular, both their critical
magnetic, $B_0$, and electric, $E_0$, fields (in the absence each time
of the other field) reach a vanishing value at the critical
temperature $T_c$.

Consider now the nanoscopic limit, namely when $a\ll\lambda(T),\xi(T)$.
In practice, this situation may be encountered indeed for nanoscopic
samples close to the critical temperature $T_c$. In such a case,
one may develop series expansion solutions in $u$ for the functions $j(u)$
and $f(u)$ in order to evaluate the criticality conditions 
(\ref{eq:critical1}) and (\ref{eq:critical2}). However, since
contributions of order $b^2$ and $e^2$ appear on both sides of these
equations, in order to be of sufficient accuracy,
the expansion in $u$ must at the same time include at least
the first order corrections in $b^2$ and $e^2$ in the r.h.s. of
(\ref{eq:critical1}) and (\ref{eq:critical2}) as well. 
For this reason, it is more relevant to consider a weak field
expansion for the solutions independently of whether $u_a$ is small or not, 
to be used to compute to first order in $b^2$ and $e^2$ the r.h.s. of 
the criticality conditions above, and then 
eventually take the nanoscopic limit. Note that given the result 
(\ref{eq:macro1}), critical fields are at least of the order of 
$\kappa/\sqrt{2}$, so that such a weak field expansion should be
warranted for small values of the LG parameter $\kappa$, namely 
for type I superconductors.

After some work, one then finds that the criticality conditions
(\ref{eq:critical1}) and (\ref{eq:critical2}),
evaluated to first order in $b^2$ and $e^2$, imply the following
constraint on the physical fields in the $(B,E)$ phase diagram,
\begin{equation}
\left(\frac{B}{B_0}\right)^2+C\left(\frac{E/c}{B_0}\right)^2\simeq 1\ ,
\label{eq:nano1}
\end{equation}
where as before $B_0$ stands for the critical magnetic field value
in the absence of any electric field, $E=0$, which is in general a function
of temperature and of $a$ of course, while $C$ is a factor
given by the following expressions,
\begin{equation}
\begin{array}{r c l}
{\rm covariant\ model}&:&\ \ C=\left(\frac{1+\beta}{1-\beta}\right)\ ,\\
 & & \\
{\rm noncovariant\ model}&:&\ \ 
C=\left(\frac{u_a}{u_a-\tanh u_a}\right)
\left(\frac{1}{1-\beta}\right)\ ,
\end{array}
\end{equation}
where
\begin{displaymath}
\beta=\frac{u_a}{16(u_a-\tanh u_a)^2}\frac{1}{(\kappa^2-2)^2}\left\{
8\kappa\sqrt{2}\frac{\tanh^2u_a}{\tanh(\kappa\sqrt{2}u_a)}-
(3\kappa^4-10\kappa^2+16)\tanh u_a+\right.
\end{displaymath}
\begin{equation}
\left.+(5\kappa^4-22\kappa^2+16)\tanh^3u_a
+(\kappa^2-2)(3\kappa^2-4)\frac{u_a}{\cosh^4u_a}\right\}\ .
\label{eq:B}
\end{equation}

These expressions are valid in the weak field approximation to
first order whatever the value for $a$. Taking now the nanoscopic limit 
as well, one finds $\left(1-\tanh(u_a)/u_a\right)^{-1}=
3/u^2_a\left[1+{\cal O}(u^2_a)\right]$
and $\beta=1/2\left[1+{\cal O}(u^2_a)\right]$, leading finally to the
following criticality conditions in the $(B,E)$ phase diagram
in the weak field limit,
\begin{equation}
\begin{array}{r c l c l}
{\rm covariant\ model}&:&\ \ 
\left(\frac{B}{B_0}\right)^2+3\left(\frac{E/c}{B_0}\right)^2\simeq 1\ &,&
\ a\ll\lambda(T),\xi(T)\ ,\\
 & & & & \\
{\rm noncovariant\ model}&:&\ \ 
\left(\frac{B}{B_0}\right)^2+6\left(\frac{\lambda(0)}{a}\right)^2
\frac{1}{1-\left(\frac{T}{T_c}\right)^4}\left(\frac{E/c}{B_0}\right)^2\simeq 
1\ &,& \ a\ll\lambda(T),\xi(T)\ .
\label{eq:nano2}
\end{array}
\end{equation}
Since the critical magnetic field $B_0$ does vanish towards the
critical temperature $T=T_c$, so do all the critical fields $B$ and $E$
which are defined by either of these relations, and thus in particular also
the critical electric field $E_0$ in the absence of any magnetic field,
$B=0$, as was already remarked previously in the noncovariant case. 
However, by having chosen to normalize the measurements of these fields 
to $B_0$, a very distinctive feature appears for the covariant model when
compared to the noncovariant one. Indeed, the ratio $E/(cB_0)$ always retains
a finite and nonvanishing value, whatever the critical values for $B$ and
$E$ within the intervals $[0,B_0]$ and $[0,E_0]$, {\sl even in the limit 
of the critical temperature $T_c$\/}, whereas in the noncovariant model,
that same ratio $E/(cB_0)$ must vanish like $\sqrt{1-(T/T_c)^4}$ (given our
chosen model for $\lambda(T)$). In particular, in the weak field
approximation and including the result (\ref{eq:macro1}) valid for 
macroscopic samples, one thus derives in the covariant case the following
bounds for the critical electric field $E_0$,
\begin{equation}
\sqrt{\frac{1-\beta}{1+\beta}}<\frac{E_0/c}{B_0}<1\ ,
\label{eq:boundE0cov1}
\end{equation}
with $E_0/(cB_0)$ moving towards lower values within that interval
when the critical temperature $T_c$ is approached (recall that $\beta$
is also temperature dependent through $u_a$). In the nanoscopic limit 
$a\ll\lambda(T),\xi(T)$, these same bounds reduce to
\begin{equation}
\frac{1}{\sqrt{3}}<\frac{E_0/c}{B_0}<1\ .
\label{eq:boundE0cov2}
\end{equation}
In contradistinction in the noncovariant case, the lower bound 
on $E_0/(cB_0)$ always vanishes, since one then finds,
\begin{equation}
\sqrt{\left(1-\frac{1}{u_a}\tanh u_a\right)(1-\beta)}<\frac{E_0/c}{B_0}<1\ ,
\label{eq:boundE0non1}
\end{equation}
reducing in the nanoscopic limit $a\ll\lambda(T),\xi(T)$ to
\begin{equation}
\frac{1}{\sqrt{6}}\left(\frac{a}{\lambda(0)}\right)
\sqrt{1-\left(\frac{T}{T_c}\right)^4}<\frac{E_0/c}{B_0}<1\ .
\label{eq:boundE0non2}
\end{equation}

As a matter of fact, this type of consideration may be refined further
still in the covariant case. Indeed, an obvious solution to the covariant LG
equations is $j(u)=\sinh u/\cosh u_a$, $f(u)=1$ in the case that
$e_{\rm ext}=b_{\rm ext}$, a fact which, as was remarked previously, is
a distinctive feature of the covariant approach, since this solution
defines precisely the boundary between the magnetic and electric regimes
of superconductivity, and as such its existence is a direct consequence
of Lorentz covariance. Hence, the critical condition 
(\ref{eq:critical1}) for the corresponding fields $b_1$ and $e_1$ 
simplifies in this specific instance to the exact result, valid under all
circumstances,
\begin{equation}
b_1^2+e_1^2=2b_1^2=2e_1^2=\frac{1}{2}\kappa^2
\frac{1}{1-\frac{1}{u_a}\tanh u_a}\ .
\end{equation}
When the weak field approximation is also warranted for the evaluation
of the critical magnetic field $B_0$, this result combines
with those above to lead to the following bounds,
\begin{equation}
\frac{1}{\sqrt{2}}\sqrt{1-\beta}<\frac{B_1}{B_0}=\frac{E_1/c}{B_0}<
\frac{1}{\sqrt{2}}\ ,
\label{eq:boundE1cov1}
\end{equation}
and in the nanoscopic limit,
\begin{equation}
\frac{1}{2}<\frac{B_1}{B_0}=\frac{E_1/c}{B_0}<\frac{1}{\sqrt{2}}\ \ \ ,
\ \ \ a\ll\lambda(T),\xi(T)\ ,
\label{eq:boundE1cov2}
\end{equation}
whereas in the noncovariant case, one finds similarly in the
nanoscopic limit
\begin{equation}
\frac{1}{\sqrt{1+6\left(\frac{\lambda(0)}{a}\right)^2
\frac{1}{1-\left(\frac{T}{T_c}\right)^4}}}<
\frac{B_1}{B_0}=\frac{E_1/c}{B_0}<\frac{1}{\sqrt{2}}\ \ \ ,\ \ \ 
\ \ \ a\ll\lambda(T),\xi(T)\ .
\label{eq:boundE1non2}
\end{equation}
Hence here again for those specific configurations such that 
$B_{\rm ext}=E_{\rm ext}/c$, the lower bound on $B_1/B_0$
reaches a vanishing value at the critical temperature in the
noncovariant case, whereas that lower bound remains finite and is only
mildly temperature dependent in the covariant case.

The existence of such finite bounds on the values for $E_1/(cB_1)$ 
as a function of temperature in the covariant case, translates into
the following nice characterization in terms of the $(B/B_0,E/(cB_0)$ phase
diagram. Indeed, the limits (\ref{eq:boundE1cov2}) (which are more
refined in (\ref{eq:boundE1cov1})) imply that the phase boundary
curve in that diagram must always cross the diagonal line $B=E/c$ within the 
interval of $B/B_0$ or $E/(cB_0)$ values defined by these bounds in 
the covariant model, whatever the value for the temperature (see Fig.1). 
Such a property is simply not met in the noncovariant model (see Fig.2). 
Similarly, the lower bounds 
(\ref{eq:boundE0cov1}) or (\ref{eq:boundE0cov2}) on $E_0/(cB_0)$ imply that, 
when approaching the critical temperature, the same phase boundary curve 
at $B/B_0=0$ cannot move below a specific finite value in the covariant model, 
while it must necessarily do so in the noncovariant one.

As a conclusion thus, which should remain valid beyond the
specific limits con\-si\-de\-red here, it appears that by choosing 
to normalize the measurement of critical electric and magnetic fields 
to the critical magnetic field in the absence of any electric field, 
for a given nanoscopic sample with this specific geometry of applied
fields and by approaching the critical temperature,
the $(B,E)$ phase diagram provides the necessary distinctive features
which should enable to discriminate experimentally between the covariant
and noncovariant mo\-dels, and in any case confirm or invalidate the
description offered by the Lorentz covariant LG equations. Indeed, as
was remarked previously, the ordinary noncovariant framework is not
physically realistic when it comes to nanoscopic samples in the presence
of electric fields, since it ignores the partial penetration, albeit
small, of the electric field into the sample's surface.
The present analysis has concentrated on the weak field approximation,
essentially in the nanoscopic limit. Similar distinctive differences
between the two models should also exist for larger values of $\kappa$,
in ways still to be investigated requiring then a detailed numerical 
study which is not pursued in this Letter.

\vspace{10pt}

\noindent{\bf 5. Numerical solutions.}
Here, we present the results of the numerical resolution of the
LG equations and of the criticality conditions 
(\ref{eq:critical1}) and (\ref{eq:critical2}) for only
one situation, which is close enough both to the discussion of the
previous section and to an experimentally realistic situation.
Namely, we take the following parameter values
\begin{equation}
\frac{a}{\lambda(0)}=5\ \ \ ,\ \ \ \kappa=0.02\ .
\label{eq:values}
\end{equation}
Indeed, this value for $\kappa$ is typical for aluminium (Al),
while tabulated values of $\lambda(0)$ 
for Al---$\lambda(0)=16-50$ nm with $T_c=1.18$ K---would imply that
the slab is then a few hundred nanometers thick,
within reach of present lithographic techniques for Al on
a SiO$_2$ substrate. Moreover, the critical magnetic field $B^\infty_c(0)$
for Al is also on the order of 100 Gauss, so that the required
electric field values for a measurement of the $(B,E)$ phase
diagram would reach into 3 MV/m, namely 3 V/$\mu$m, certainly
also a reasonable range of values for such a nanoscopic device.
Of course, compared to the infinite slab model, such a device will
be subjected to finite size corrections. Presumably, such corrections
would imply that the role played by $\lambda(T)$ and $\kappa$ in our
analysis would be replaced by some effective quantities whose values
would not differ to a great extent from those of Al in the bulk.
Such corrections may be assessed only once a specific device is designed.

In Fig.1 (resp. Fig.2), we present the 
$(B/B_0,E/(cB_0))$ phase diagram for the covariant (resp. noncovariant) 
model, given the values in (\ref{eq:values}), for a series of temperatures 
in the range from $T=0$ to $T=T_c$. The general behaviour
of the phase diagram as a function of temperature is indeed the one described
in the previous section. In particular in the covariant model,
and as a function of temperature, the critical electric field values 
$E_0/(cB_0)$ and $E_1/(cB_0)$ obey the different 
finite lower (and upper bounds) derived from the analytical discussion,
including those given in (\ref{eq:boundE0cov1}) and (\ref{eq:boundE1cov1})
when considering the associated values for $\beta$. In contradistinction,
in the noncovariant case, the ratio $E/(cB_0)$ reaches a
vanishing value when approaching the critical temperature, while in this
case as well it way be checked that the different lower bounds 
(\ref{eq:boundE0non1}), (\ref{eq:boundE0non2}) and
(\ref{eq:boundE1non2}) are indeed also obeyed.

Such results, as well as the other considerations of this Letter
show that it should be possible to experimentally 
discriminate between the ordinary noncovariant LG equations and 
the covariant ones advocated here, by determining the critical
$(B,E)$ phase diagram of a nanoscopic superconducting sample for 
temperatures approaching its critical temperature. 
The geometry of the applied fields is crucial
for this purpose, with the external magnetic field parallel to the
sample's surface and the external electric field perpendicular to it.
By normalizing the measurement of fields to that of the critical magnetic
field in the absence of any electric field, distinctive differences
between the two approaches are best brought to the fore, and should
enable to confirm or invalidate the covariant approach. Moreover,
if the experiment should also allow for an absolute calibration
of the applied fields, the comparison between the two models may be
refined still further by considering the temperature dependency of the
critical value for applied magnetic $B$ and electric $E/c$ fields of
equal magnitude, this temperature dependency being
constrained to lie within a specific interval
whose existence is a direct consequence of the manifest Lorentz covariance
of the covariant model. We hope to be able to report on such measurements in 
the future, but lithographic problems have hindered any progress until now.

\noindent{\bf Acknowledgements.}
J.G. wishes to thank Peter van Nieuwenhuizen and the members of the
C.N. Yang Institute for Theoretical Physics at the State University of
New York at Stony Brook for their kind hospitality. The work
of G.S. is financially supported as a Scientific Collaborator of the
``Fonds National de la Recherche Scientifique" (FNRS, Belgium).

\clearpage

\noindent{\bf Figure Captions}

\vspace{10pt}

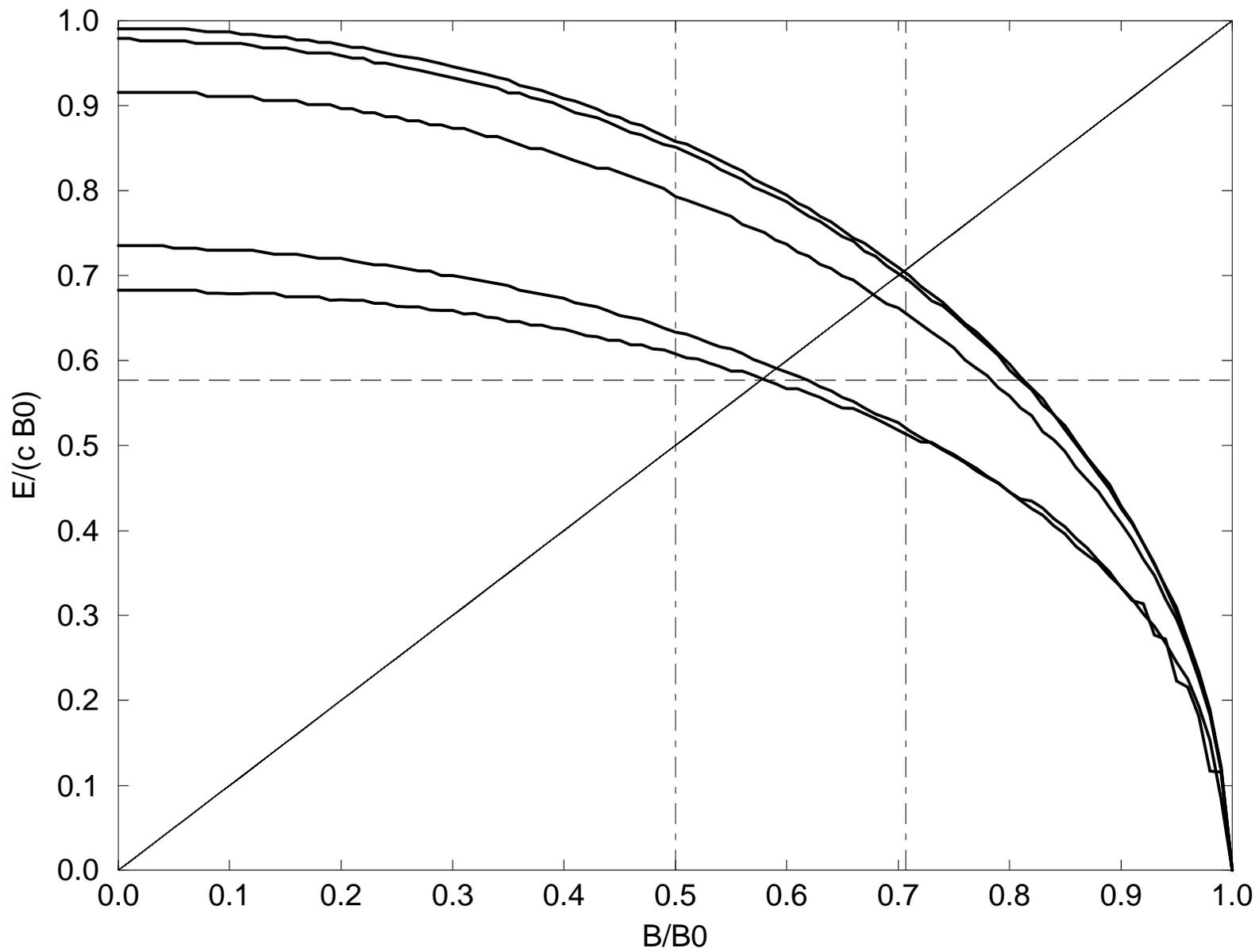
\noindent Figure 1: The phase diagram $(B/B_0,E/(cB_0))$ for the
covariant LG equations with the va\-lues (\ref{eq:values}). Shown
from top to bottom are the curves associated to the following
increasing temperature values, $T/T_c=0,0.8766,0.9659,0.9935,0.9996$.
The diagonal line determines those configurations such that
$B_{\rm ext}=E_{\rm ext}/c$, the two vertical dot-dashed lines
at $B/B_0=1/2$ and $B/B_0=1/\sqrt{2}$ correspond to the lower and 
upper bounds (\ref{eq:boundE1cov2}) obeyed by the critical electric $E_1/c$ 
and magnetic $B_1$ fields of equal strength in the nanoscopic limit
of the weak field ap\-pro\-xi\-mation for the covariant LG equations,
while the horizontal dashed line at $E/(cB_0)=1/\sqrt{3}$ corresponds to
the lower bound (\ref{eq:boundE0cov2}) on the critical electric field
$E_0/(cB_0)$ in the same approximation.
The existence of these finite bounds is the distinctive prediction
of the covariant model and a direct consequence of its manifest Lorentz
covariance.

\vspace{20pt}

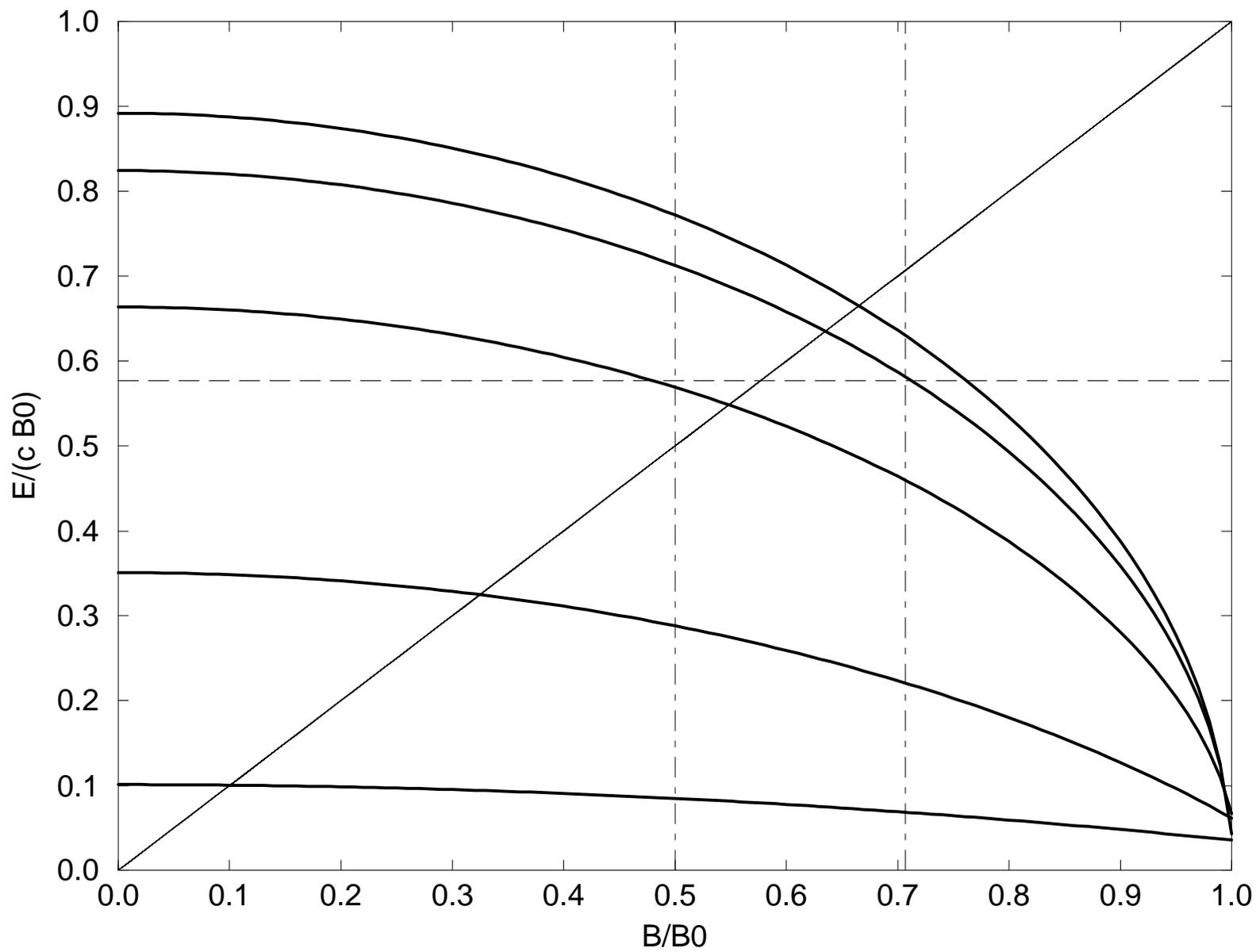
\noindent Figure 2: The same as in Fig.1 for the noncovariant
model. In this case, the horizontal and two vertical lines are displayed 
only for the purpose of comparison with the covariant model.


\clearpage

\newpage

\begin{thebibliography}{99}

\bibitem{Tinkham} See for example,\\
M. Tinkham, {\sl Introduction to Superconductivity\/}, 2nd Edition
(McGraw Hill, New York, 1996);\\
J.R. Waldram, {\sl Superconductivity of Metals and Cuprates\/}
(Institute of Physics, Bristol, 1996).

\bibitem{Josephson} B.D. Josephson, {\sl Phys. Lett.\/} {\bf 16} (1965) 242.

\bibitem{Bok} J. Bok and J. Klein, {\sl Phys. Rev. Lett.\/}
{\bf 20} (1968) 660.

\bibitem{Kolacek} J. Kol\'a\u{c}ek and E. Kawate, {\sl Phys. Lett.\/}
{\bf A 260} (1999) 300;\\
J. Kol\'a\u{c}ek and P. Va\u{s}ek, {\sl Hall voltage
sign reversal in type II superconductors\/}, {\tt cond-mat/9911222};\\
J. Kol\'a\u{c}ek, P. Lipavsk\'y and V. \u{S}pi\u{c}ka,
{\sl Electric field in type II superconductors\/}, {\tt cond-mat/9911283}.

\bibitem{Weinberg} S. Weinberg, {\sl The Quantum Theory of Fields\/}
(Cambridge University Press, Cambridge, 1996), Vol. II, pp. 332-352.

\bibitem{Gov1} J. Govaerts, G. Stenuit, D. Bertrand and O. van der Aa,
{\sl Phys. Lett.\/} {\bf A 267 } (2000) 56 ({\tt cond-mat/9908451}).

\bibitem{Gov2} J. Govaerts, in preparation.

\end{thebibliography}
\end{document}